\documentclass[12pt]{article}
\usepackage{arxiv}

\usepackage{natbib}

\usepackage{hyperref}       
\usepackage{url}            
\usepackage{amsfonts}       
\usepackage{comment}

\RequirePackage{amsthm,amsmath,amsfonts,amssymb}
\usepackage{placeins}
\usepackage{float}
\RequirePackage{graphicx}
\graphicspath{{Figures/}}

\usepackage{algorithm}
\usepackage{algpseudocode}
\usepackage{bbm,bm}
\usepackage{mathtools}
\usepackage{multirow}

\title{Time-varying $\ell_0$ optimization for Spike Inference from Multi-Trial Calcium Recordings}

\author{
  Tong Shen\\
  Department of Statistics\\
  University of California\\
  Irvine, CA 92697
  \And
  Kevin Johnston\\
  Department of Mathematics and \\
  The NSF-Simons Center for Multiscale Cell Fate Research\\
  University of California\\
  Irvine, CA 92697

  \And 
  Gyorgy Lur\\
  Department of Neurobiology and Behavior\\
  University of California, Irvine.

  \And
  Michele Guindani\\
  Department of Statistics\\
  University of California\\
  Irvine, CA 92697
  
  \And
  Hernando Ombao\\
  Statistics Program, KAUST, Thuwal, SA.
  \And
    Xiangmin Xu\\
   Department of Anatomy and Neurobiology\\
   School of Medicine\\
   University of California\\
  Irvine, CA 92697

   \And
 Zhaoxia Yu \\
  Department of Statistics\\
  University of California\\
  Irvine, CA 92697\\
  \texttt{zhaoxia@ics.uci.edu} 
}

\begin{document}

\maketitle

\begin{abstract}
Optical imaging of genetically encoded calcium indicators is a powerful tool to record the activity of a large number of neurons simultaneously over a long period of time from freely behaving animals. However, determining the exact time at which a neuron spikes and estimating the underlying firing rate from calcium fluorescence data remains challenging, especially for calcium imaging data obtained from a longitudinal study. We propose a multi-trial time-varying $\ell_0$ penalized method to jointly detect spikes and estimate firing rates by robustly integrating evolving neural dynamics across trials. Our simulation study shows that the proposed method performs well in both spike detection and firing rate estimation. We demonstrate the usefulness of our method on calcium fluorescence trace data from two studies, with the first study showing differential firing rate functions between two behaviors and the second study showing evolving firing rate function across trials due to learning.  
\end{abstract}


\section{Introduction}
In the past few years, calcium imaging has been increasingly adopted in neuroscience research since it allows simultaneous measurement of the activity of a large population of neurons at the single-neuron resolution over weeks using optical imaging of living animals. For example, we recently conducted a longitudinal investigation of the neural ensemble dynamics of contextual discrimination by recording mice's calcium imaging in the hippocampus for about 60 days  \citep{johnston2020robust} using  appropriate genetically encoded calcium indicators \citep{tian2009imaging} and miniature fluorescence miscroscopes \citep{ghosh2011miniaturized}. The technical advancements bring great flexibility to neuroscience research; they also create significant challenges to every aspect of data analysis - from storing the large amount of video recordings to  downstream statistical modeling and inference \citep{pnevmatikakis2019analysis}. In longitudinal recordings of freely behaving animals, after motion correction and registration of detected neurons across multiple sessions \citep{giovannucci2019caiman}, fluorescence traces of individual neurons can be extracted, for example using independent component analysis \citep{mukamel2009automated} or nonnegative matrix factorization methods \citep{maruyama2014detecting, pnevmatikakis2016simultaneous,zhou2018efficient}. Strategies to reduce false positive rates of neuron detection have also been proposed, such as our recent work on imposing spatial constraints on footprints of neurons \citep{johnston2020robust}. The extracted fluorescence traces of individual neurons allow for the statistical analysis of complex neural interactions at the single cell level. Fluorescence traces are  often used for visualization, clustering neurons based on similar neural activity profiles, comparing neural activity levels between various experimental conditions, and studying neural encoding and decoding. 

However, calcium fluorescence data provide a noisy proxy, rather than direct observations of neural activity. For many important questions, such as those involving the analysis of the precise timing of neural activity in response to stimuli, it is essential to estimate the underlying spike train from a noisy fluorescence trace. Several approaches have been developed including linear deconvolution \citep{yaksi2006reconstruction} or nonnegative convolutions \citep{vogelstein2010fast}. Fully Bayesian methods have also been developed to obtain posterior distributions of spike trains, thus allowing uncertainty quantification of spikes' estimates \citep{pnevmatikakis2013bayesian}. 
\citep{theis2016benchmarking} proposed a supervised learning approach based on a probabilistic relationship between fluorescence and spikes. This algorithm is trained on  data where spike times are known. 
Research has shown that nonnegative deconvolution outperforms supervised algorithms and is more robust to the shape of calcium fluorescence responses \citep{pachitariu2018robustness}. One limitation of these methods is that these spike detection methods analyze only one trial at time; thus, shared information across trials is largely ignored. In a longitudinal study, neural activities are measured in multiple trials or sessions. In these settings, aggregating information across trials could increase the accuracy of spike detection.

There has been limited work on how to utilize the information from multiple trials to improve accuracy in spike detection. Among the few multi-trial methods we identified, \cite{picardo2016population} assumed that repeated trials share the same burst time but have trial-specific magnitude, baseline fluorescence, and noise level. Conceptually, integrating multiple trials should be beneficial if the trial-to-trial variation is mainly due to randomness. In reality, however, as pointed out in \cite{deneux2016accurate}, the gain by naively combining trials may be limited - it is likely to bring improvement for some trials but might perform worse in others when the neural activity in some trials does not follow the marginal pattern across all trials. During a learning process or when adjusting to a new environment, neurons constantly reorganize and show plasticity, which leads to varying neural dynamics across trials. 

In this paper, we develop a robust multi-trial spike inference method to address the challenges from longitudinal calcium imaging data. Our approach both integrates the commonality and accounts for evolving neural dynamics across trials. Efficiency is achieved by aggregating information from temporally adjacent trials whereas robustness is guaranteed by incorporating a time-varying firing rate function into a dynamic $\ell_0$ penalization framework, rather than forcing shared parameters across trials.

\section{Methods}
In this section, we first propose a time-varying $\ell_0$ penalized framework to analyze multi-trial calcium trace data. We then present details on how to implement it by alternating a firing rate estimation step and a spike detection step. 

\subsection{Multi-Trial time-varying $\ell_0$ penalized auto-regressive model (MTV-PAR)}
Let $y_r(t)$ be the fluorescence recorded at time point $t$ of trial $r$, where $t=1,...T$, $r=1,\cdots,R$. In practice, multiple pre-processing steps, including normalization,  \citep{vogelstein2010fast}, are typically implemented to obtain $y_r(t)$. The calcium fluorescence trace $y_t(t), t=1,\cdots, T$ is often modeled using the following first-order auto-regressive model

\begin{equation}
\begin{split}
     y_r(t) &= c_r(t)+\epsilon_r(t), \>\>\> \epsilon_r(t)\sim N(0,\sigma_r^2), \\
     c_r(t) &= \gamma_r c_r(t-1)+s_r(t),  \text{ s.b.t. } s_r(t)\ge 0
     \end{split}
\label{eqn:ar1}
\end{equation}

\noindent where $c_r(t)$ and $\epsilon_r(t)$ denote the underlying calcium concentration and noise at time $t$ in the $r$th trial, respectively; $0<\gamma_r(t)<1$ is the decay parameter for a calcium transient, which depends on multiple factors such as sampling frame, cell types, and the kinetics of genetically encoded indicators. In this model, $s_r(t)$ denotes the change in calcium concentration between time $t-1$ and time $t$ with $s_r(t)>0$ indicates a spike occurs at time $t$ and 0 means no spike.
The main goal of spike detection to locate the time points with a positive $s_r(t)$. 

Here, we propose to regularize the inference on spike detection by introducing a time-varying penalty function $\lambda_r(t)$ on the number of spikes, as the spikes tend to be sparse but not evenly distributed over time:  

\begin{equation}
\min_{c_r(1), \ldots, c_r(T)} \left\{\frac{1}{2} \sum_{t=1}^{T}(y_r(t)-c_r(t))^{2}+ \sum_{t=2}^{T}\lambda_r(t) \mathbbm{1}_{s_r(t) > 0}\right\}.
\label{eqn:l0_opt1}
\end{equation}

\noindent We further assume that $\lambda_r(t)$ is a decreasing function of the instantaneous firing rate $f_r(t)$, which is assumed to be smooth over time $t$ within trials and change slowly over trials $r$. Because the parameters in the AR(1) model are trial-specific but the firing rate functions across trials are assumed to change slowly to capture the evolving neural dynamics, efficiency and robustness are well balanced. In addition, our modeling approach allows the simultaneous estimation of spike trains and firing rate functions. 

In the rest of this subsection, we present an iterative approach that (1) estimates the firing rate function, given the spike trains derived in the previous iterations, by using a local nonparametric method; (2) detects spike trains, given the current estimate of the firing rate function, by using a time-varying $\ell_0$ penalization.

\subsection{Firing rate estimation and time-varying penalty}
Estimating the firing rate function is a crucial task  in the analysis of spike train data \citep{cunningham2009methods}. One commonly used descriptive approach involves building  the peristimulus time histogram (PSTH) where the spike counts are  averaged from multiple trials within each time bin \citep{gerstein1960approach}. Kernel methods are often applied  to achieve smoothness   \citep{cunningham2008inferring}. Bayesian methods have also been considered, e.g. by proposing Gaussian processes \citep{cunningham2008inferring} and Bayesian Adaptive Regression Splines \citep{dimatteo2001bayesian, olson2000neuronal, kass2003statistical, kass2005statistical, behseta2005testing}. 

The PSTH approach and the other smoothing methods  assume that the underlying firing rate function does not change over trials. The estimated firing rates are then compared between different groups to capture the association between firing rate and animal behavior \citep{jog1999building,wise1999role,wirth2003single}. However, recent evidence suggests to move beyond the independent and identically distributed trial assumption and regard both neural and behavioral behavioral dynamics as smooth and continuous \citep{huk2018beyond, ombao2018statistical}. Thus, it is essential to model the between-trial dynamics. To account for the between-trial dynamics, some authors have proposed a state-space framework  \citep{czanner2008analysis,paninski2010new}. In the state-space framework, the spike train is characterized by a point process model \citep{brown2003likelihood, brown2005theory, kass2005statistical, daley2007introduction} for the underlying fire rate. In this paper, to make it feasible to  jointly estimate the spike trains and the firing rate function from the observed multi-trial fluorescence data,  we consider instead a computationally less demanding two-dimensional Gaussian-boxcar kernel smoothing function $G(r,t)$, which is formulated as follows
\begin{equation}
G(r,t)=\frac{1}{\sqrt{2 \pi} \sigma} \exp \left(-\frac{t^{2}}{2 \sigma^{2}}\right) I(|r|<B/2)    
\label{eqn:Gaussian-boxcar}
\end{equation}
\noindent where $\sigma$ denotes the within-trial kernel bandwidth in the Gaussian kernel, $I(\cdot)$ is the indicator function, and $B$ is a bandwidth for the between-trial sliding windows in the boxcar kernel. Thus, our estimate of $f_r(t)$ is given by 
$$
\hat f_r(t) \propto \sum_{r',t'}  G(r-r',t-t') p_{r'}(t'),
$$
where $p_{r'}(t')$ is the estimated number of spikes per second in a small time bin centered at time $t'$ in trial $r'$.

For PSTH based on spike train data, common choices of the Gaussian bandwidth are around 50-150 ms \citep{cunningham2009methods}. When choosing the optimal bandwidth to smooth a spike train estimated from  calcium fluorescnce trace data, one should also consider the fluorescence transient kinetics. \cite{ali2019interpreting} reviewed the kinetics of calcium transients with several fluorescent indicators. For example, they reported that GCaMP6f has a rise time of 42 ms and decay time of 142 ms while GCaMP6s has a rise time of 179 ms and decay time of 550 ms; a time-bin of 200 ms is suggested in one of their analyses.  In our analysis, $\sigma$ is chosen between 200ms and 400ms. For the box-car bandwidth to allow borrowing information across trials, we use a pre-selected window size $B$, which is mainly determined by to the number of available trials. As discussed in the Discussion section, data driven methods, such as the one used in \citep{fiecas2016modeling, ombao2018statistical}, will be considered in future work.

In spike detection, sparsity has been enforced via penalization or introducing appropriate prior distributions. Existing approaches usually adopt a tuning parameter uniformly for the whole time series of a fluorescence trace. Within a trial, the firing rate right after each stimulus are expected to be higher than baseline. Thus,  using a constant penalty may not be optimal. Our proposed non-constant penalization is inspired by prior work in the literature. 
For example, time-varying penalization was used for analyzing multivariate time series data \citep{fan2013composite, yu2017frm}.  \cite{zbonakova2016time} studied the dynamics of the penalty term in a Lasso framework for the analysis of interdependences in the stock markets. \cite{monti2017learning} used varying regularization for different edges in a Gaussian graphical model to study brain connectivity in fMRI data.

Here, we expand on those approaches and consider  a decreasing function of the firing rate function for the time-varying penalty, motivated by the fact that spikes are expected to be less frequent in regions with low firing rates than in  regions with high firing rates. Ideally, the penalty should be small in the locations with higher firing rates. Hence, we  use a negative exponential function \citep{tang2010reliability} to achieve  adaptive regularization. Specifically, the time-varying function is chosen as
$$\lambda_r(t) \propto e^{-a\hat f_r(t)}$$
where $\hat f_r(t)$ is the estimated firing rate (See Algorithm 2). Here the value $a$ controls how much the penalty function should depend on the firing rate function. In particular, $a=0$ reverts to a constant penalty. To avoid extremely large or small penalties, we scale the estimated firing rate function of each trial by dividing by its maximum value, so that it ranges between 0 and 1. This implies that the default value $a=1$ in our analysis leads to mildly time-varying penalties - within a trial, the penalty at the highest firing rate is about $37\%$ of the penalty at a firing rate of $0$. In addition, to facilitate comparison with the case of a constant penalty, the following formula is used in each trial to scale the penalty function $\lambda_r(t)$ to have mean $\lambda$,
\begin{equation}
    \lambda_{r}(t) = \lambda \, T \times \,  \frac{\exp\left\{-a\frac{\hat f_{r}(t)}{\max(\hat f_{r}(t))}\right\}}  
    {\sum_{t'=1}^T \exp\left\{-a\frac{\hat f_{r}(t')}{\max(\hat f_{r}(t'))}\right\}}.  
\end{equation}

\subsection{Time-varying penalized $\ell_0$ AR(1) model}
In our proposed multi-trial time-varying $\ell_0$ penalized auto-regressive model (MTV-PAR), a calcium fluorescence trace at the $r$th trial $y_r(t), t=1,\cdots, T$, is modeled by a first-order auto-regressive model, as given in Equation (\ref{eqn:ar1}). As previously stated,  $c_r(t)$ denotes the underlying calcium concentration and a positive value of $s_r(t)$ implies that a spike occurs at time $t$. 

Because spikes tend to be sparse, ideally, one should penalize the number of spikes by introducing an $\ell_0$ penalty. However, $\ell_0$ penalization is computationally intractable; therefore, existing methods often impose an $\ell_1$ penalization \citep{vogelstein2010fast,friedrich2016fast,friedrich2017fast}. Recently, \cite{jewell2018exact} found that, for spike detection, the use of an $\ell_0$ penalty brings substantial improvements over an $\ell_1$ penalty. They also showed that relaxing the nonnegative constraint of $s_r(t)$ has a negligible effect on the results but the corresponding $\ell_0$ optimization problem is equivalent to a change point detection problem whose solution can be obtained efficiently using a dynamic programming algorithm. Use a similar strategy, we prove that the following time-varying penalization problem can also be solved by a dynamic programming algorithm: 
\begin{equation}
\min_{c_r(1), \ldots, c_r(T)} \left\{\frac{1}{2} \sum_{t=1}^{T}(y_r(t)-c_r(t))^{2}+ \sum_{t=2}^{T}\lambda_r(t) \mathbbm{1}_{s_r(t) \not= 0}\right\}.
\label{eqn:l0_opt2}
\end{equation}

Specifically, we find that the time-varying $\ell_0$ optimization problem is equivalent to the following change point problem whose solution can be efficiently identified using a dynamic programming algorithm. For the ease of presentation, we drop the trial index $r$ and focus on the fluorescence trace of a cell at a given trial. The equivalent change point problem can be shown as (See Appendix \ref{appA}.1 for proof): 
\begin{equation}
\min_{0=\tau_{0}<\tau_{1}<\ldots<\tau_{k}<\tau_{k+1}=T, k}\left\{\sum_{j=0}^{k}\left[ \mathcal{D}\left(y_{\left(\tau_{j}+1\right): \tau_{j+1}}\right)+\lambda(\tau_j)\right] \right\},
\end{equation}
where $\tau_1,...\tau_k$ are $k$ change points, i.e., the points satisfying $c(t)-\gamma c(t-1) \neq 0$ and 
\begin{equation}
\mathcal{D}(y_{a: b}) = \min _{\substack{c(a), c(t)=\gamma c(t-1)\\t=a+1, \ldots, b}}\left\{\frac{1}{2} \sum_{t=a}^{b}(y(t)-c(t))^{2}\right\},
\end{equation}

\noindent which has the following closed-form solution: 
\begin{equation}\mathcal{D}(y_{a:b})=\frac{1}{2} \sum_{t=a}^{b}\left(y(t)-\gamma^{t-a} \frac{\sum_{t=a}^{b} y(t) \gamma^{t-a}}{\sum_{t=a}^{b} \gamma^{2(t-a)}}\right)^{2}.
\end{equation}

\noindent Note that the parameter $\gamma$, which measures the speed at which the calcium concentration decays, is not estimated. The value of $\gamma$  is usually close to 1 \citep{vogelstein2010fast, yaksi2006reconstruction}, as a somatic calcium transient caused by an action potential is often characterized by an almost instantaneous rise but a slow decay. For computational feasibility, rather than estimating $\gamma$ iteratively, similar to \citep{pnevmatikakis2016simultaneous, friedrich2017fast}, we estimate it using the auto-correlation at lag 1.

Finally, as shown in Appendix \ref{appA}.2, the optimization problem (\ref{eqn:l0_opt2})
can be solved by computing $F(t)$ recursively: \begin{equation}
F(t)=\min\limits_{0 \leq \tau<t}\left\{F(\tau)+\mathcal{D}\left(y_{(\tau+1): t}\right)+\lambda(\tau)\right\}, \>\>\>\mbox{for } t=1,2,\cdots, T. 
\end{equation}
	
\noindent The resulting algorithm has a time complexity $\mathcal{O}(n^2)$. This optimization problem can be solved in $\mathcal{O}(n)$ time (Algorithm 1) with a dynamic programming algorithm \citep{auger1989algorithms, jackson2005algorithm, jewell2018exact, killick2012optimal} . 

\begin{algorithm}[H]
		\caption{Dynamic programming algorithm to detect spikes with a time-varying penalization function}
		\label{euclid}
		\begin{algorithmic}[1]
			\State \textbf{Input:} Time-varying penalty function $\lambda(t)$, single-trial fluorescence $y(t)$ with $T$ time points.
			\State\textbf {Initialize: } $F(0)=-\lambda, spikeset=\emptyset, \mathcal{E}_{1}=\{0\}$
			\For{$t = 1,2,...T$}
		    \State Calculate $F(t)=\min\limits_{\tau\in \mathcal{E}_t}\left\{F(\tau)+\mathcal{D}\left(y_{(\tau+1): t}\right)+\lambda(\tau)\right\}$
		    \State Find $t^\ast=\underset{\tau\in\mathcal{E}_s}{\operatorname{argmin}}\{F(\tau)+\mathcal{D}\left(y_{(\tau+1): t}\right)+\lambda(\tau)\}$
		    \State Let $\mathcal{E}_{t+1}=\left\{\tau \in\left\{\mathcal{E}_{t} \cup t\right\}: F(\tau)+\mathcal{D}\left(y_{(\tau+1): t}\right)<F(t)\right\}$
		    \If{$t^\ast \notin$ $spikeset$ }
		    \State Add $t^\ast$ to $spikeset$
		    \EndIf
			\EndFor
			\State Return $Spikeset$
		\end{algorithmic}
	\end{algorithm}

\subsection{Algorithms}
Thus far, we have presented our solutions to two problems separately. The first problem is to use multi-trial spike data to estimate the firing rate function $f_r(t)$ for $r=1,\cdots, R$ and $t=1,\cdots,T$ using a Gaussian-boxcar smoothing method. The second problem is to detect spikes from a single calcium fluorescence trace using a time-varying $\ell_0$ penalized approach under the assumption that the time-varying penalty function $\lambda_r(t)$ is already known. In practice, neither firing rate functions nor spike locations are known. We therefore propose an Expectation-Maximization-type algorithm to jointly estimate firing rate and detect spike locations. As shown in Algorithm 2, we alternate between the spike detection and firing rate estimation steps until the spike indicator function does not change anymore.

		\begin{algorithm}[H]
		\caption{Simultaneous spike detection and firing rate estimation}
		\begin{algorithmic}[1]
			\State\textbf{Input:} Multi-trial fluorescence data $y(t)^{R\times T}$ with $R$ trials and $T$ time points. Penalty term $\lambda$.
			\State\textbf{Initialize}: Penalty function $\lambda_{r}(t)=\lambda$.
			\State Set the binary spike indicator $x(r,t)^{R\times T} = 0$. 
			 
			\While{ the indicator matrix $x(r,t)^{R\times T}$ changes}
			\For{$r = 1,2,...R$} 
		      \State Apply Algorithm 1 to detect spikes and let $spikeset$ denote the times at which spikes were detected. 
		      \State Set $x(r,spikeset) = 1.$
		    
			\EndFor
			
			  \State Estimate the firing rate $\hat{f}_{r}(t)$, for $r=1,\cdots,R$ and $t=1,\cdots, T$ using Gaussian-Sliding-Window kernel smoothing.
			  \State Calculate the weight function $$w_{r}(t) = e^{-a\frac{\hat f_{r}(t)}{\max(\hat f_{r}(t))}}$$
			  \State Update the time-varying penalty function $\lambda_{r}(t) =  \lambda*T*w_{r}(t)/\sum_{t'} w_{r}(t')$ 
			\EndWhile
		\end{algorithmic}
	\end{algorithm}
	

    
\section{Simulation}
\subsection{Metrics for quantifying accuracy}
We conducted a set of simulation studies to evaluate the performance of our MTV-PAR method. Among the many competing methods, we choose the $\ell_0$ approach in \cite{jewell2018exact} because its performance has been shown better than competing methods using both simulated and benchmark data. When comparing estimated spike trains with the ground truth, we use the Victor-Purpura (VP) distance \citep{victor1996nature,victor1997metric}, which has been commonly used for comparing spike trains. It is defined as the minimum total cost required to transform one spike train into another using the following three basic operations:
    \begin{itemize}
    \item Insert a spike into a spike train. (Cost = 1)
    \item Delete a spike. (Cost = 1)
    \item Shift a spike by an interval $\Delta t$. (Cost = $q\Delta t$) A large $q$ makes the distance more sensitive to fine timing differences. We use the default value $q=1$.
    \end{itemize}
    
Because MTV-PAR estimates the firing rate function together with the spikes, we also evaluate its performance on firing rate estimation. The approach in \cite{jewell2018exact} does not estimate firing rates; hence, for a fair comparison, we apply the same Gaussian-boxcar kernel smoothing in (\ref{eqn:Gaussian-boxcar}) to the spikes estimated using \cite{jewell2018exact} in order to estimate the  firing rates. 
The accuracy of firing rate estimation is calibrated by the $\ell_2$ norm  \citep{adams2009tractable} of the difference between an estimated firing rate function and the true function: 
    \begin{equation*}
    \|f(t)-\hat{f}(t)\|_2 = \left(\frac{1}{\int m(t)dt} \int|f(t)-\hat{f}(t)|^2 m(t)dt\right)^{1/2}
    \end{equation*}
    where $m(t)$ is the weight at $t$ and its default value is a constant.  

 \subsection{Simulation of spike trains from inhomogeneous Poisson processes}
 
 In this simulation setting, the trials are treated as repeated trials. In other words, the trials share the same underlying firing rate function and each trial is an independent realization of an inhomogeneous Poisson process. Several forms of firing rate functions have been considered in previous work. For example, \cite{kass2003statistical, behseta2005testing} assumed  bell-shaped firing rate functions. \cite{pachitariu2018robustness} used a piecewise constant stimulus rate. Firing rate functions with multiple peaks from exponential stimuli functions have also been considered \citep{reynaud2014goodness}. In our simulation, we chose the following bi-modal firing rate function
	\begin{equation}
	f(t) = 0.01 + 0.19*\sum\limits_{t_0}\exp(-(t-t_0)^2/d^2) \>\>\> (t=1,2,...1000)
	\label{eqn:firing_rate_constant}
	\end{equation}
where $d=150$ and the stimuli peaks are at $t_0=300$ and $700$. Thus,  $f(t)$ reaches its maximum value 0.2 (equivalent to 10 spikes per second) at time 300 and 700. 

To generate spike trains for multiple trials under an inhomogeneous Poisson process with the firing rate function in \ref{eqn:firing_rate_constant}, we followed the idea of \cite{adams2009tractable}. In each trial, the spikes were first randomly drawn from a homogeneous Poisson process. A thinning process was then applied to create a realization from the desired inhomogeneous Poisson process (see Appendix \ref{appB} for details). After obtaining the simulated spike trains, we generated calcium fluorescence traces using the auto-regressive model \eqref{eqn:ar1}. The following parameters were used in the simulations:  $T=1000$, $\gamma = 0.96$, $\sigma = 0.15$ and $R=50$ trials in total. According to \cite{vogelstein2010fast}, the parameters  above correspond to a sampling rate of 50 Hz and a length of 20 seconds per trial. For each simulation setting, 100 data sets were generated.
 
\begin{figure}[H]
		\centering
		\begin{tabular}{c c}
	\includegraphics[scale=0.4]{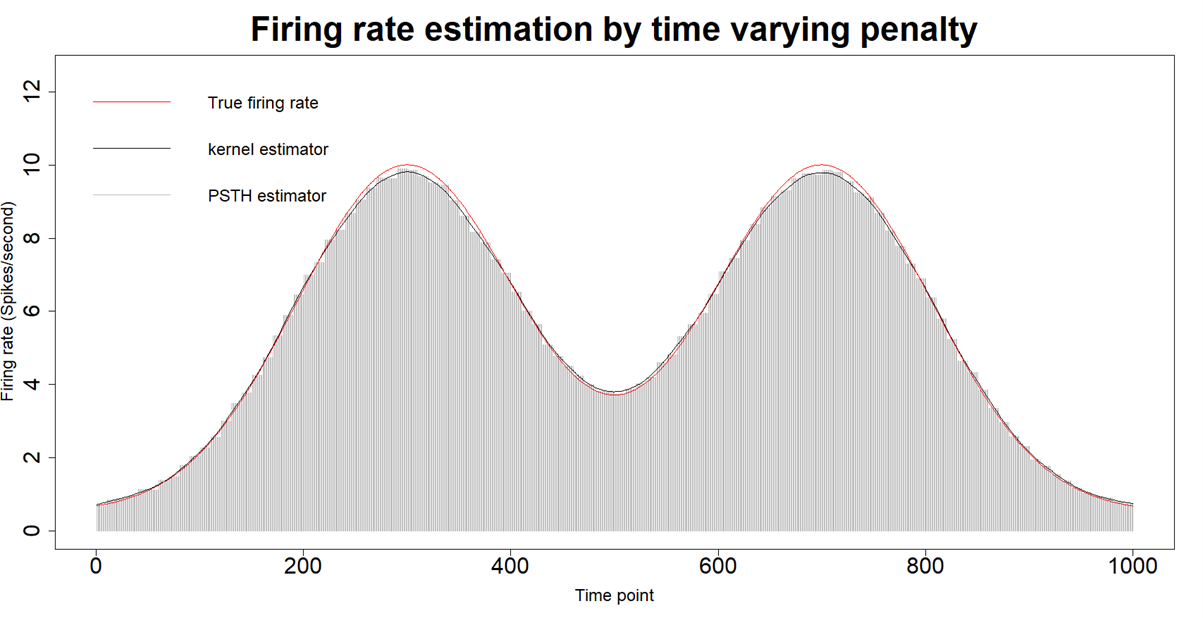}&\includegraphics[scale=0.4]{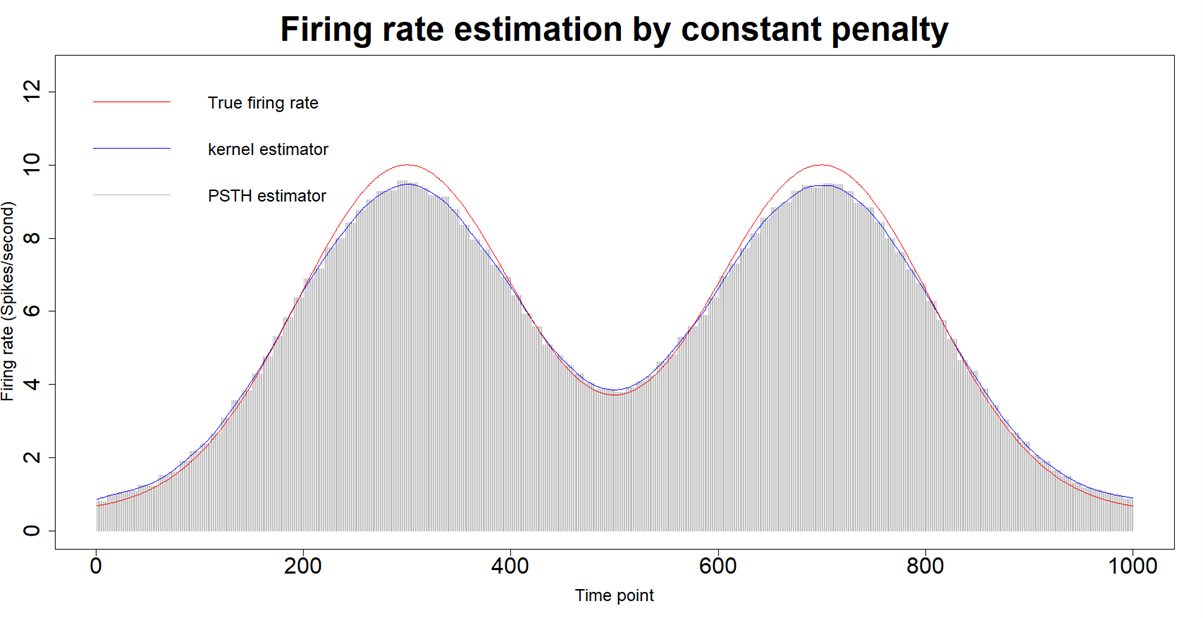}\\
	\includegraphics[scale=0.6]{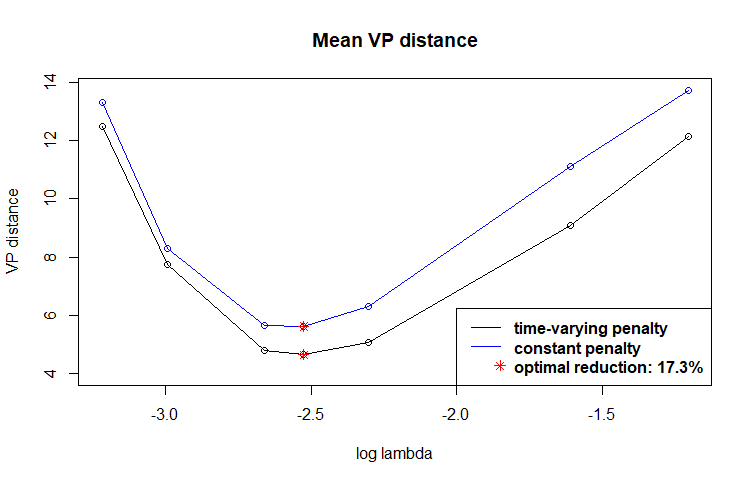}&
	\includegraphics[scale=0.6]{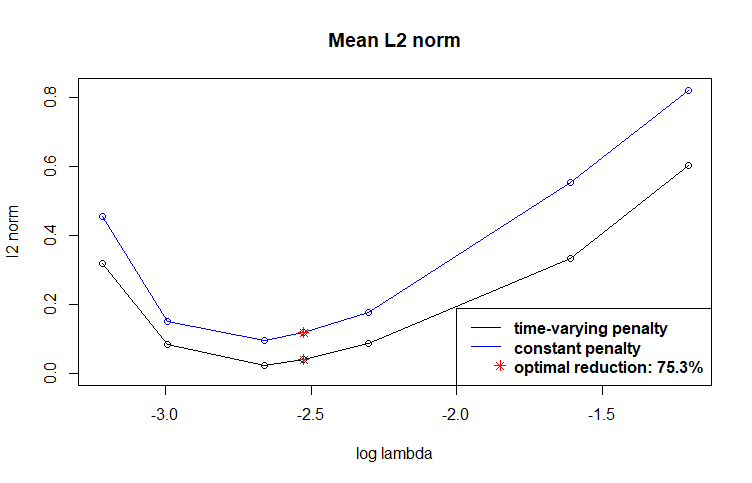}
		\end{tabular}
		\caption{Top: firing rate estimates using MTV-PAR (time-varying) and constant penalty. Bottom: VP distance of spike trains and $\ell_2$ norm of firing rates between the truth and estimators.}
		\label{fig:sim_constant}
	\end{figure}

 We implemented our MTV-PAR using Algorithm 2. When estimating $\lambda(t)$, we chose the Gaussian kernel bandwidth equal to 200 ms. Because the trials are assumed to have the same underlying firing rate function,
 B=50 is used, which is equivalent to averaging the spike counts across trials to estimate the firing rate $f(t)$. As shown in the top panel of Figure \ref{fig:sim_constant}, the firing rate functions estimated from MTV-PAR are closer to the true underlying function. Thus, adopting the proposed  time-varying penalty led to an improvement in both spike detection (measured by mean VP distance, the lower panel of Figure \ref{fig:sim_constant}) and firing rate estimation (measured by the $\ell_2$ norm, the upper panel of Figure \ref{fig:sim_constant}) across all the $\lambda$ values considered. For the optimal $\lambda$ based on VP, the reductions brought by the  time-varying penalty in VP and $\ell_2$ norm are $17.3\%$ and $76.3\%$, respectively. 

We also simulated data with a larger noise level ($\sigma=0.3$), longer series ($T=2,000$) and higher decay rate ($\gamma = 0.98$). In all scenarios, we observed improved accuracy from using the proposed MTV-PAR (results omitted).
	
\subsection{Simulation under a dynamic firing rate function}
\begin{figure}[H]
		\centering
		\begin{tabular}{c c}
			\includegraphics[scale=0.65]{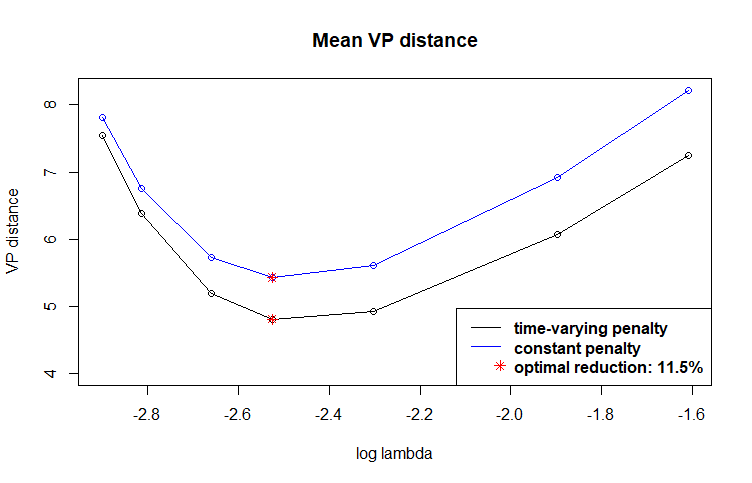}&\includegraphics[scale=0.65]{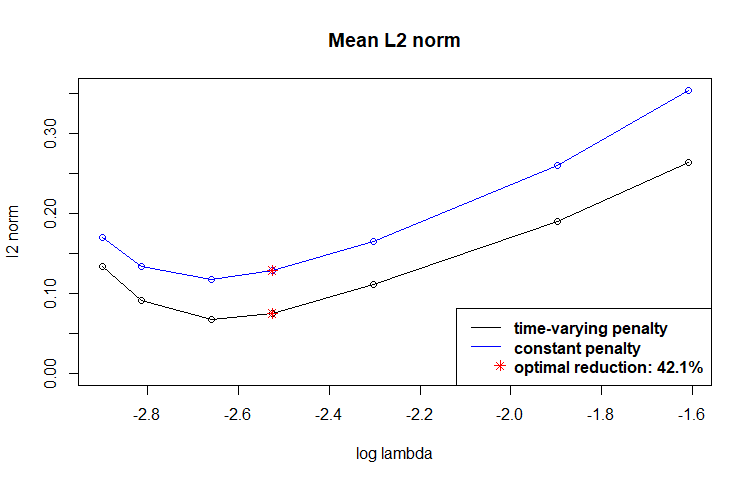}
		\end{tabular}
		\caption{Top Left: True firing rates; Top Middle: Estimated firing rates using MTV-PAR; Top Right: Estimated firing rates using constant penalty. Bottom: VP distance of spike trains and L2 norm of firing rates between the truth and estimators. }
		\label{fig:sim_dynamic}
	\end{figure}

We also simulated data under dynamic firing rate functions across trials. Figure \ref{fig:sim_dynamic} (the upper left panel) shows a dynamic firing rate function with two peaks within each trial; across trials, the peak values first increase then decrease. The two-dimensional firing rate function at time point $t$ and trial $r$ is as follows:
	\begin{equation}
	f_{r}(t) = 0.01 + 0.19\left(\exp\left\{-\frac{(t-300)^2}{150^2}\right\}+ \exp\left\{-\frac{(t-700)^2}{150^2}\right\}\right)*\exp\left\{-\frac{(r-R/2)^2}{1000}\right\}
	\end{equation}
where $t=1,2,...,T$ and $r=1,2,...,R$. 

Figure \ref{fig:sim_dynamic} summarizes the simulation results with $T=1,000$, $\gamma = 0.96$, $\sigma = 0.15$ and $R=50$. The sampling rate is 50 Hz and the length is 20 seconds per trial. The maximum of the firing rate is 0.2 (equivalent to 10 spikes per second). The results are based on 100 simulations. In firing rate estimation, Gaussian kernel smoothing is applied with a kernel bandwidth of 200 ms within trials and a window length of 10 across trials. Similar to the simulation in Section 3.2, MTV-PAR performs much better than using a constant penalty in estimating spikes and firing rate. Specifically, at the optimal $\lambda$ (based on minimal VP distance), the reduction brought by MTV-PAR in VP distance for spike detection is $11.5\%$ and the reduction of $\ell_2$ error of firing rate estimation is $42.1\%$.

	\begin{figure}[H]
		\centering
		\begin{tabular}{c c c}
			\includegraphics[scale=0.53]{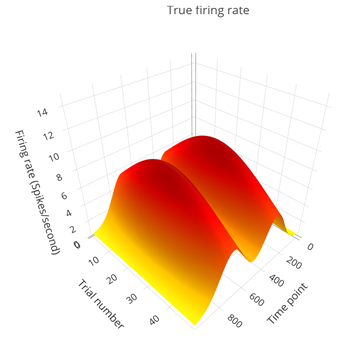}&
			\includegraphics[scale=0.5]{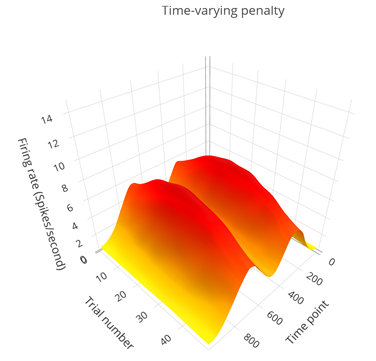}&\includegraphics[scale=0.5]{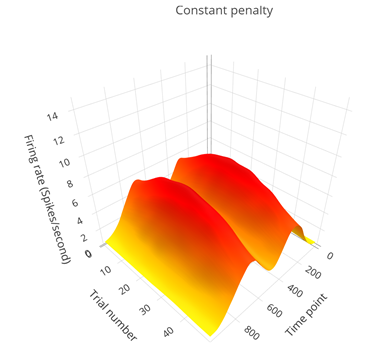}
		\end{tabular}
	\end{figure}
	
\section{Application to Calcium Imaging Data}
We now pursue the scientific investigations on the neural activities in two longitudinal mice studies in which both calcium imaging data and behavioral data were recorded. In this first data set, calcium imaging data were collected within a few days after the participating mice had been trained for a discrimination task; thus, it is reasonable to treat trials with same behavior outcome as repeated trials. As a comparison, in the second study, calcium recording started in the first learning trial and lasted for a few weeks; therefore, neural dynamics are expected as a result of the learning process and neural plasticity. For this reason, incorporating neuronal dynamics in firing rate estimation is likely to improve spike detection and firing rate estimation. 

\subsection{Mouse task data I}
The activity of neurons (labelled with GCaMP6s) from the mouse's anterior lateral motor cortex (ALM) was recorded with two-photon calcium imaging during a head-fixed whisker-based discrimination\ task \citep{li2015motor}. In the experiment, mice were supposed to discriminate the pole locations using their whiskers and report the perceived pole position by licking. Each trial is composed of three epochs: sample epoch (mice presented with a vertical pole), delay epoch (the pole was removed), response epoch (mice cued to give a response). If a mouse licked the correct lick port, it was rewarded with liquid.

\begin{figure}[H]
		\centering
		\begin{tabular}{c c}
			\includegraphics[scale=0.6]{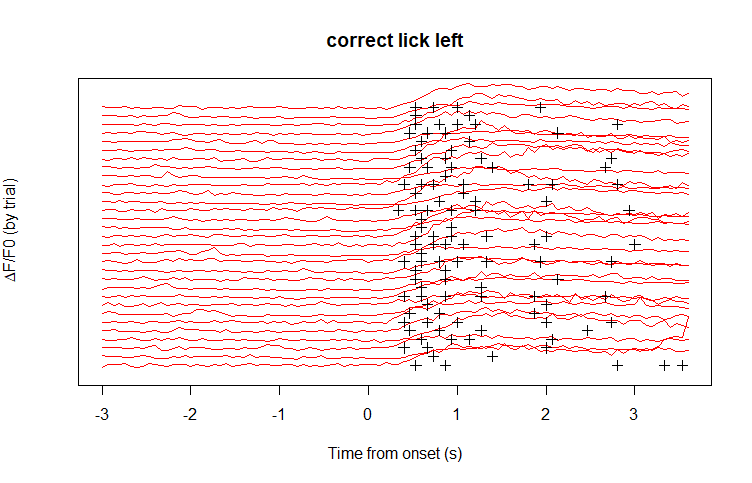}&\includegraphics[scale=0.6]{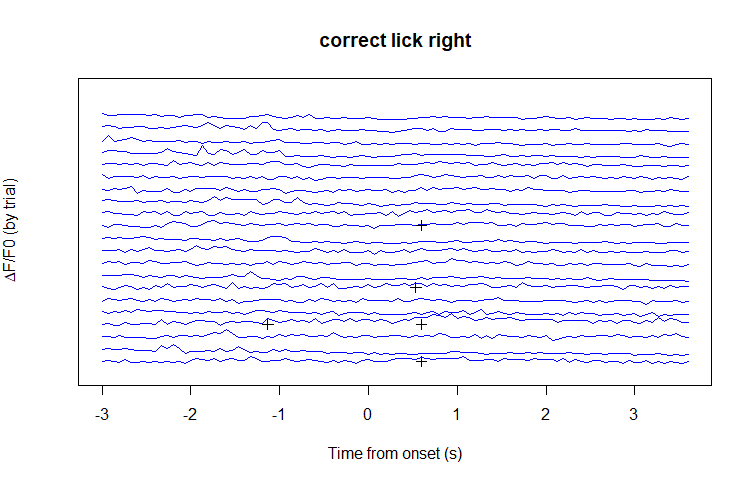}\\
			\includegraphics[scale=0.6]{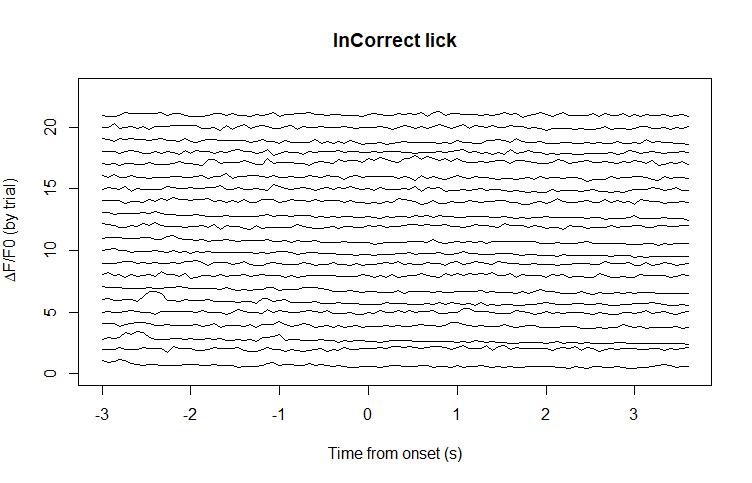}&\includegraphics[scale=0.6]{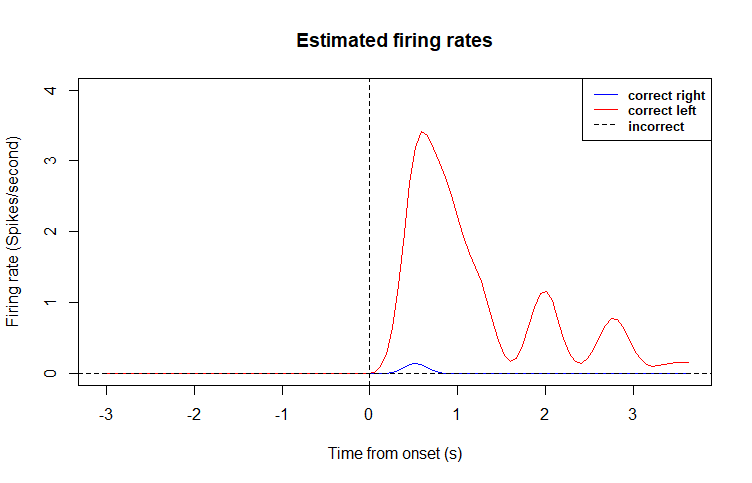}
		\end{tabular}
		\caption{An example cell from the water lick experiment \citep{li2015motor}. Top Left: The calcium fluorescence trace under correctly lick left outcome (black dots: estimated spikes). Top Right: The calcium fluorescence trace under correctly lick right outcome. Bottom Left: The calcium fluorescence trace under incorrect lick outcomes. Bottom Right: Estimated firing rate functions of under different conditions. Vertical dashed line: the start of response epoch.}
		\label{fig:water_lick}
	\end{figure}

The data we present here is from one mouse with 73 trials from multiple days. Each trial contained more than 100 data points at 15 Hz. The fluorescence data were obtained after typical pre-processing procedures such as correction for neuropil contamination and  $\Delta F/F_0$ transformation where $F_0$ is the baseline averaged fluorescence within a $0.5$s period right before the start of each trial. There were four possible behavioral outcomes in the experiment: correct/incorrect lick left/right. Because most of the trials were either ``correct lick left" or ``correct lick right", we combined the two incorrect groups as a single group. Figure \ref{fig:water_lick} shows the calcium fluorescence traces of a pyramidal tract neuron in 73 trials, including 31 trials of ``correct lick left", 21 trials of ``correct lick right", and 21 trials of ``incorrect lick" . 

One interesting question is whether the neuron responded differently for different outcomes. Therefore, we conducted a stratified analysis for each outcome. In the firing rate estimation, we use a Gaussian kernel smoothing with bandwidth 200 ms within trials. Since the mouse has been well trained before calcium imaging recording, firing rates across trials under the same outcome group are relatively stable, which was confirmed by available 2D visualization (data not shown). Thus, it is sensible to combine all the trials within an outcome type when estimating firing rate. 

As indicated in Figure \ref{fig:water_lick}, in ``correct lick left" trials, the neuron fired right after the cue time (when the mouse was cued to make decision); however, there was almost no neural activity under the other two outcome groups. The estimated firing rate function also confirmed this difference. It is known that the ALM brain region of mice is involved in planning directed licking \citep{guo2014flow}. The estimated spikes and firing rate functions provide convincing evidences that this neuron is likely to show neural selectivity and play a critical role in the cognitive process of making the correct decision of licking the left pole.

\subsection{Mouse task data II}
\begin{figure}[H]
		\centering
		\begin{tabular}{c c}
			\includegraphics[scale=0.3]{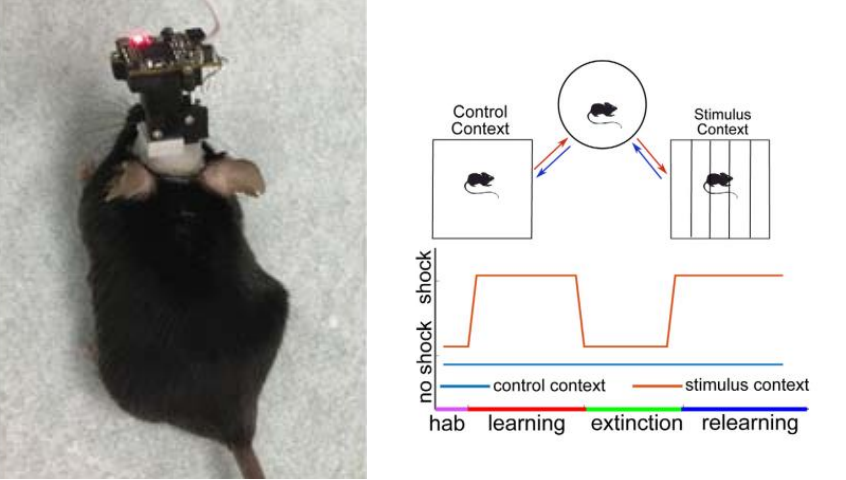}&\includegraphics[scale=0.75]{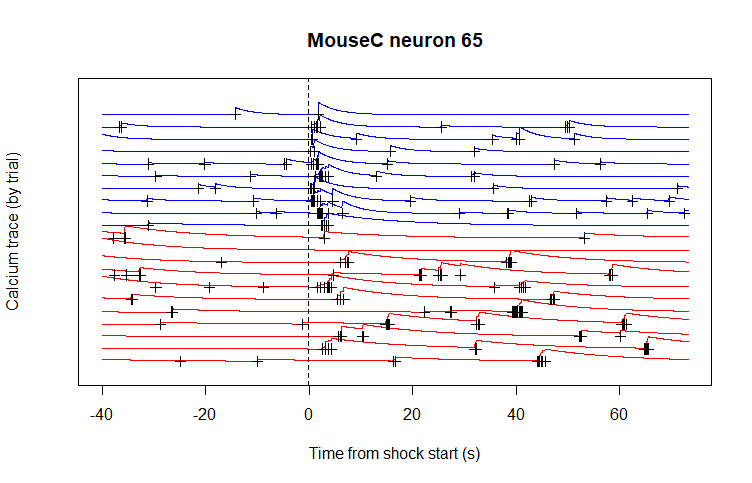}\\
			\includegraphics[scale=0.4]{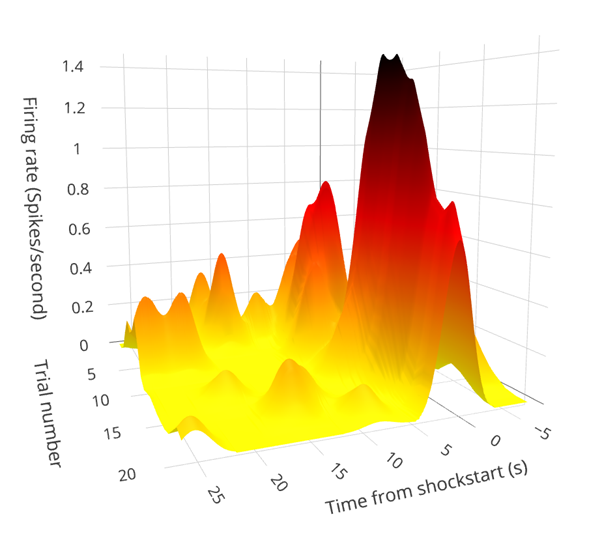}&\includegraphics[scale=0.75]{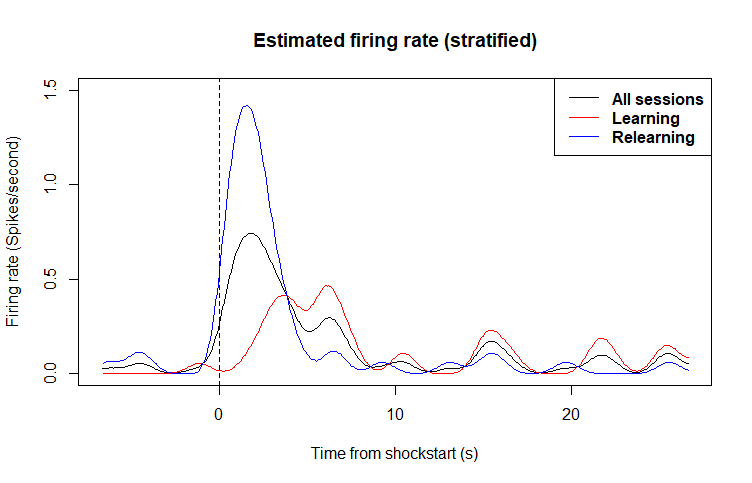}
		\end{tabular}
		\caption{An example neuron from the foot shock study. Top Left: Calcium imaging of mouse hippocampus using a miniaturized scope and the four stages of the experimental design: habituation, learning, extinction, and relearning \citep{johnston2020robust}. Top Right: Calcium fluorescence traces and estimated spikes of a sample neuron. The red and blue traces were obtained during the learning and relearning stages, respectively. The long dashed vertical line denotes the time when a foot shock was applied and the short black vertical lines are the time locations at which spikes were detected. Bottom Left: the estimated firing rate function using MTV-PAR. Bottom Right: the estimated firing rate function using stratified analysis. Vertical dashed line: the start time of foot shock.}
		\label{fig:foot_shock}
	\end{figure}

The second data set is from our study of long-term contextual discrimination experiment, in which mice were trained to recognize two contexts via fear conditioning (foot shock) \citep{johnston2020robust}. The research goal was to understand the behavior-associated hippocampal neural ensemble dynamics at single-cell resolution. Viral injections were administered into mice brain to introduce a genetically encoded calcium indicator (GCaMP6f).  Fluorescence signals from hippocampal CA1 excitatory neurons were then optically recorded using one photon head-mounted miniscopes from freely moving mice (Figure \ref{fig:foot_shock}, upper left). There were four stages in the experiment: habituation (mice freely exploring environment), learning (learning to freeze in a stimulus context with foot shock), extinction (no foot shock), and relearning (stimulus reinstated). 

In the mouse we analyzed, the activity of 141 neurons were recorded for several weeks. We chose the 21 foot shock sessions with the first 11 sessions in the learning stage and the remaining 10 sessions in the relearning stage. In each shock session, a foot shock was administered and the fluorescence  trace was recorded at 15 Hz for 2 minutes. Figure \ref{fig:foot_shock} (upper right panel) shows an example neuron. The calcium fluorescence  traces from different trials were temporally aligned by the start of shock time. For the firing rate estimation, we apply the Gaussian-boxcar kernel smoothing with a bandwidth of 400 ms for within trials and a window length of 5 for across trials. The estimated spikes (Figure \ref{fig:foot_shock}, upper right) and firing rate functions (Figure \ref{fig:foot_shock}, lower left) suggest that the neuron is more synchronous to the stimulus in the relearning stage than in the learning stage, which may reflect the evolving learning-related neuronal dynamics.

It is worth noting that treating the trials as independent realizations of the same underlying process will lead to undesired results. As shown in Figure \ref{fig:foot_shock} (lower right panel), when assuming the trials as samples from the \underline{same} underlying distribution, the estimated peak firing time is misleading. Although the stratified estimates from the two stages showed that neuronal firing took place sooner and was more frequent in the relearning stage than in the learning stage in response to the foot shock stimulus, the results based on stratified analysis were not able to completely characterize the intrinsic evolution of neural firing during the cognitive learning process. 

\section{Discussion}
In this paper, we proposed MTV-PAR, a time-varying $\ell_0$ penalized method to simultaneously conduct spike detection and firing rate estimation from longitudinal calcium fluorescence trace data. In MTV-PAR, each iteration consists of two steps: the spike detection step using time-varying $\ell_0$ regularization based on the current estimate of the firing rate functions and the firing rate estimation step based on the currently detected spikes. Our approach is able to account for the intrinsic neural dynamics at both within- and between-trial levels. 
Here we used a time-varying $\ell_0$ penalization in the spike detection step and a Gaussian-boxcar smoothing in the firing rate estimation step. In some situations, other types of regularization and smoothing methods might be preferred or more appropriate; our strategy of integrating information from multiple trials can be adopted in a similar manner.

Our work was motivated by recovering the underlying spike data from noisy calcium imaging and the main focus had been the accuracy of spike detection. Borrowing information across neighboring trials improves spike estimation; as a consequence, the firing rate estimation is also improved. In our local nonparametric estimation for firing rate functions, for computational ease, we chose the window size based on experience and the total number of available trials. The window size and other tuning parameters are best calibrated using benchmark data. In the absence of benchmark data, data driven methods might be helpful. \cite{cunningham2007inferring} proposed to use a Gaussian process prior and an iterative gradient optimization algorithm to optimally choose hyperparameters. \cite{fiecas2016modeling} proposed to use the window size that minimizes the generalized cross-validated deviance when investigating the evolution of dynamic interactions (coherence) between brain regions. Adopting similar strategies is likely to gain further insights and efficiency when analyzing calcium imaging data.

Like other existing methods, we aim to estimate the time of a neural spike or a burst of spike events. Bounded by the temporal resolution of calcium imaging data, when there are multiple spikes within the same time frame, we are not able to zoom in to estimate the times of individual spikes. In several works, the change in fluorescence intensities due to spike events at time $t$, i.e., $s(t)$,  was initially motivated to model spike counts \citep{vogelstein2010fast, friedrich2016fast, friedrich2017fast}; following this line of thought, it is reasonable to assume that $s(t)$ is positively correlated with the spike counts for re-scaled fluorescence traces. In other work \citep{pnevmatikakis2013bayesian, jewell2018exact}, the sign of $s(t)$ was the focus. The unknown and nonlinear relationship between spike counts and fluorescence transient measured based upon genetically encoded indicators \citep{vogelstein2010fast, lutcke2013inference, rose2014putting} complicates the accurate interpretation and modeling of fluorescence transient data. In the current version of MTV-PAR, the magnitude of $s(t)$ was not used in the firing rate estimation. It is worth conducting future research to examine how to best utilize $s(t)$ in analyzing calcium fluorescence data.

\newpage
\begin{appendix}
\section{Proofs for Section 2.3}\label{appA}
\subsection{Equivalence between Problem (5) and Problem (6)}
\begin{equation*}
\begin{aligned}
 &\min_{c(1), \ldots, c(T)}\left\{\frac{1}{2} \sum_{t=1}^{T}(y(t)-c(t))^2+ \sum_{t=2}^{T}\lambda(t) 1_{(c(t)-\gamma c(t-1) \neq 0)}\right\}\\
 &\Leftrightarrow\min_{\substack{0=\tau_{0}<\tau_{1}<\ldots<\tau_{k}<\tau_{k+1}=T, k\\ c(1),...,c(T)}}\left\{\sum_{j=0}^{k}\frac{1}{2}\sum_{t=\tau_j+1}^{\tau_{j+1}}(y(t)-c(t))^2 +\sum_{t=2}^{T}\lambda(t) 1_{c(t)-\gamma c(t-1) \neq 0} \right\}\\
 &\Leftrightarrow\min_{\substack{0=\tau_{0}<\tau_{1}<\ldots<\tau_{k}<\tau_{k+1}=T, k\\c(1),...,c(T)}}\left\{\sum_{j=0}^{k}\frac{1}{2}\sum_{t=\tau_j+1}^{\tau_{j+1}}(y(t)-c(t))^2 + \sum_{j=0}^k \lambda(\tau_j) \right\}-\lambda(0)\\
 &\Leftrightarrow\min_{\substack{0=\tau_{0}<\tau_{1}<\ldots<\tau_{k}<\tau_{k+1}=T, k}}\left\{\sum_{j=0}^{k}\frac{1}{2}\sum_{\substack{t=\tau_j+1...\tau_{j+1}\\(c(t)=\gamma c(t-1))}} (y(t)-c(t))^2 + \sum_{j=0}^k \lambda(\tau_j) \right\}\\
 &\Leftrightarrow\min_{0=\tau_{0}<\tau_{1}<\ldots<\tau_{k}<\tau_{k+1}=T, k}\left\{\sum_{j=0}^{k}\left[ \mathcal{D}\left(y_{\left(\tau_{j}+1\right): \tau_{j+1}}\right)+\lambda(\tau_j)\right] \right\}
\end{aligned}
\end{equation*}

\subsection{Equivalence between Problem (5) and Problem (9)}
Based on the result in Appendix A.1, to show that Problems (5) and (9) are equivalent, we only need to show that Problems (6) and (9) are equivalent.   

\begin{equation*}
\begin{aligned}
 &\min _{0=\tau_{0}<\tau_{1}<\cdots<\tau_{k}<\tau_{k+1}=t, k}\left\{\sum_{j=0}^{k}\left[ \mathcal{D}\left(y_{\left(\tau_{j}+1\right): \tau_{j+1}}\right)+\lambda(\tau_j)\right]\right\} \\
&\Leftrightarrow\min _{0=\tau_{0}<\tau_{1}<\cdots<\tau_{k}<\tau_{k+1}=t, k}\left\{\sum_{j=0}^{k-1}\left[\mathcal{D}\left(y_{\left(\tau_{j}+1\right): \tau_{j+1}}\right)+\lambda(\tau_j)\right]+\mathcal{D}\left(y_{\left(\tau_{k}+1\right): \tau_{k+1}}\right)+\lambda(\tau_k)\right\}\\
&\Leftrightarrow\min_{0<\tau_k<\tau_{k+1}=t}\{\min _{0=\tau_{0}<\tau_{1}<\cdots<\tau_{k^\prime}<\tau_{k^\prime+1}=\tau_k, k^\prime}\left\{\sum_{j=0}^{k^\prime}\left[\mathcal{D}\left(y_{\left(\tau_{j}+1\right): \tau_{j+1}}\right)+\lambda(\tau_{j})\right]\right\}\\
&+\mathcal{D}\left(y_{\left(\tau_{k}+1\right): \tau_{k+1}}\right)+\lambda(\tau_k)\}\\
&\Leftrightarrow \min _{0 \leq \tau<t}\left\{F(\tau)+\mathcal{D}\left(y_{(\tau+1): t}\right)+\lambda(\tau)\right\}
\end{aligned}
\end{equation*}

\newpage
\section{Algorithm for Simulating Spike Trains from an Inhomogeneous Poisson Process}\label{appB}

\begin{algorithm}[H]
		\caption{Simulate Spike Trains from an Inhomogeneous Poisson Process ( Modified from Algorithm 1 of \cite{adams2009tractable})}
		\begin{algorithmic}[1]
			\State\textbf {Input} Length of data $n$, upper bound of rate intensity $f^\ast$, firing rate function $f(t)$
		    \State Draw the number of spikes: $S\sim Poisson(nf^\ast)$ 
		    \State Uniformly distribute the spikes:
		    $\mathcal{E}_S=\left\{s_i\right\}_{i=1}^{S} \sim \text { Uniform }(n)$
		    \For {$i=1,2,...S$}
		    \If {$uniform(0,1)\ge\sigma\left(f(s_i)\right)$}
		    \State exclude $s_i$ from $\mathcal{E}_S$
		    \EndIf
		    \EndFor
			\State Return the set of spikes $\mathcal{E}_S$
		\end{algorithmic}
	\end{algorithm}

\end{appendix}

\newpage

\bibliography{references}  

\end{document}